\documentclass[aps,preprint,amsmath,amssymb]{revtex4-1}
\usepackage{graphicx}
\usepackage{xcolor}
\newcommand{\nn}{\nonumber}
\newcommand{\bd}{\begin{document}}
\newcommand{\ed}{\end{document}}
\newcommand{\bc}{\begin{center}}
\newcommand{\ec}{\end{center}}
\newcommand{\be}{\begin{eqnarray}}
\newcommand{\ee}{\end{eqnarray}}
\newcommand{\ba}{\begin{array}}

\newcommand{\ea}{\ed{array}}
\newcommand{\strich}[1]{#1  \! \! \slash}
\newcommand{\eqn}{\global\def\theequation}
\newcommand{\sw}{sin^2 \theta_W}
\newcommand{\fbd}{f_B}
\renewcommand{\thefootnote}{\alph{footnote}}
\newcommand{\se}{\section}
\newcommand{\sse}{\subsection}
\newcommand{\bi}{\bibitem}
\def\figcap{\section*{Figure Captions\markboth
     {FIGURECAPTIONS}{FIGURECAPTIONS}}\list
     {Figure \arabic{enumi}:\hfill}{\settowidth\labelwidth{Figure 999:}
     \leftmargin\labelwidth
     \advance\leftmargin\labelsep\usecounter{enumi}}}
\let\endfigcap\endlist \relax
\def\reflist{\section*{References\markboth
     {REFLIST}{REFLIST}}\list
     {[\arabic{enumi}]\hfill}{\settowidth\labelwidth{[999]}
     \leftmargin\labelwidth
     \advance\leftmargin\labelsep\usecounter{enumi}}}
\let\endreflist\endlist \relax

\begin{document}
\title
{\Large {\bf Some heavy vector and tensor meson decay constants in light-front quark model}
}

\author{Chao-Qiang Geng$^{1,2,3}$\footnote{geng@phys.nthu.edu.tw},
Chong-Chung Lih$^{4,3}$\footnote{cclih@phys.nthu.edu.tw}
and
Chuanhui Xia$^{1}$\footnote{chxia@cqjtu.edu.cn}
}
\affiliation{
$^1$College of Materials Science and Engineering,
Chongqing Jiaotong University, Chongqing, 400074, China\\
$^2$Department of Physics, National Tsing Hua University, Hsinchu, Taiwan 300 \\
$^3$Physics Division, National Center for Theoretical Sciences, Hsinchu, Taiwan 300\\
$^4$Department of Optometry, Shu-Zen College of Medicine and Management,
Kaohsiung Hsien,Taiwan 452
}

\date{\today}
\begin{abstract}
We  study the  decay constants ($f_M$) of the heavy
vector  ($D^{*}$, $D^{*}_{s}$,
$B^{*}$, $B^{*}_{s}$,  $B^{*}_{c}$) and  tensor ($D_{2}^{*}$, $D_{s2}^{*}$, $B^{*}_{2}$, $B^{*}_{s2}$)
 mesons in the light front quark model.
With the known pseudoscalar meson
decay constants of  $f_D$, $f_{D_s}$, $f_B$, $f_{B_s}$, and $f_{B_c}$
 as the input parameters to determine the light-front meson wave functions,
we obtain that
$f_{D^{*}, D^{*}_{s}, B^{*},B^{*}_s,B^{*}_c} =
(252.0^{+13.8}_{-11.6}$, $318.3^{+15.3}_{-12.6}$ , $201.9^{+43.2}_{-41.4}$, $244.2\pm7.0$, $473.4\pm18.2$) and
$(264.9^{+10.2}_{-9.5}$, $330.9^{+9.9}_{-9.0}$, $220.2^{+49.1}_{-46.2}$, $265.7\pm8.0$, $487.6\pm19.2$) MeV
with  Gaussian and power-law wave functions,
  respectively,
while 
$f_{D_{2}^{*},D_{s2}^{*},B^{*}_{2},B^{*}_{s2}}$=
 ($143.6^{+24.9}_{-21.8}$, $209.5^{+29.1}_{-24.2}$,
 $80.9^{+33.8}_{-27.7}$, $109.7^{+15.7}_{-15.0}$) MeV
 with only Gaussian wave functions.

\end{abstract}


\maketitle %

\se{Introduction}

Meson decay constants contain  useful information on the nonperturbative
behavior of QCD between  quarks and antiquarks inside mesons.
In addition, the determinations of these helpful
parameters can also be used to constrain the CKM mixing matrix elements in  weak mesonic decays.
In recent  years, many heavy vector and tensor mesons  have been experimentally discovered,
such as the excited states of the charmed mesons~\cite{charm}, observed by
Babar, Belle, CLEO, FOCUS and LHCb Collaborations.
Moreover,  D0~\cite{D0} and CDF~\cite{CDF}
Collaborations have confirmed the bottom  states of $B_1$(5721), $B_2$(5747),
$B_{s1}$(5830) and $B^{*}_{s2}$(5840).
In some of these hadron states,
the quantum numbers are $I(J^{P} ) = \frac{1}{2} (2^{+})$.
The investigations of these particles are clearly important
in hadron physics both theoretically and experimentally.
The recent experimental results on the parameters of these mesons
would help us to understand the meson properties and  the
non-perturbative dynamics as well as the vacuum structure of QCD.

In the literature, the decay constants of heavy vector and tensor mesons are somewhat less discussed.
In particular, compared to the scalar and pseudoscalar mesons, there are few theoretical
works devoted to the analysis of the properties
for the tensor mesons. The main purpose of this work is to examine the vector and tensor mesons
decay constants simultaneously within the framework of the light-front quark model (LFQM),
which has  been widely used
in the phenomenological study of
 meson physics.
The LFQM is a good way for solving the
nonperturbative problems of hadron physics and provides inside  information
about the internal structure of the bound state.
The meson decay constant can be described by a two-point function and regarded as one of the
simplest physical observable in the LFQM.
This framework has been applied successfully to explain various
properties of pseudoscalar and vector mesons~\cite{LF}.

The paper is organized as follows. In Sec. II, we present
the basic formalism  of the LFQM.
In Sec. III, we show our numerical results on the decay constants in the LFQM.
Our conclusions are given in Sec. IV.

\se{Formalism}

In the LFQM, a neutral meson wave function is constructed
in terms of its constituent quark $q$ and  anti-quark $\bar{Q}$ with the total momentum $p$ and spin $S$ as~\cite{lfbs},
\begin{eqnarray}
|M(p,S,S_z)\rangle &=& \int [dk_{1}][dk_{2}] 2(2\pi)^{3}\delta
^{3}(p-k_{1}-k_{2}) \nonumber \\ &\times&\sum_{\lambda
_{1}\lambda_{2}} \Phi_{M}(k_1,k_2,\lambda_1,\lambda_2)
b_{q}^{+}(k_{1},\lambda _{1}) d_{\bar{Q}}^{+}(
k_{2},\lambda _{2}) |\,0\,\rangle\,, \label{mbs}
\end{eqnarray}
where
\be
 [dk] = {dk^+d^{2}k_{\bot}\over 2(2\pi)^3 }\,,
\ee
$M$  represents for a $P$ (pseudoscalar) or $V$ (vector) or $T$ (tensor) meson,
 $\Phi_{M}$ is the wave function of the corresponding $q\bar{Q}$ and $k_{1(2)}$
($\lambda_{1(2)}$) is the on-mass shell LF momentum (helicity) of the internal quark.
In the momentum space,  $\Phi_{M}$
can be expressed as a covariant form~\cite{lfwf1,lfwf2}
\be
\Phi_{M}(x,k_{\bot })&=&\left( \frac{%
k_{1}^{+}k_{2}^{+}}{2[M_{0}^{2}-\left( m_{q}-m_{\bar{Q}} \right) ^{2}]}\right)^{%
\frac{1}{2}}\overline{u}\left( k_{1}, \lambda _{1}\right)
\Gamma v\left( k_{2},\lambda _{2}\right) \phi_{M}(x,k_{\bot}) \,,
 \nn \\
M_0^2&=&{ m_{q}^2+k_\bot^2\over x}
+ { m_{\bar{Q}}^2+k_\bot^2\over 1-x} \, ,
\label{n6}
\ee
where  $\phi_M(x,k_{\bot})$ describes the momentum distribution
amplitude of the bound state for the $S$ or $P$-wave
meson, $(x,k_{\bot})$ are LF relative momentum
variables, defined by
\be
&& k^+_1=x p^+, \quad k^+_2=(1-x) p^+\,,  \nonumber \\
&& k_{1\bot}=x p_\bot+k_\bot, \quad k_{2\bot}=(1-x)
p_\bot-k_\bot\,,
\label{lfvb}
\ee
 and $\Gamma $ stands for
\be
&&\Gamma_{P}=\gamma_5 \qquad ({\rm pseudoscalar}, S=0),
\nonumber\\
&&\Gamma_{V}=i\bigg\{\not{\!\hat{\varepsilon}}(S_z)-
          {\hat{\varepsilon}\cdot(k_1-k_2)
                \over M_0+m_q+m_{\bar{Q}}}\bigg\} \qquad ({\rm vector}, S=1),
\nonumber\\
&&\Gamma_{T}=i\frac{\hat{\varepsilon}^{\mu\nu}}{2}\bigg\{\gamma_{\mu}-
          {(k_1-k_2)_{\mu}
                \over M_0+m_q+m_{\bar{Q}}}\bigg\} (k_1-k_2)_{\nu}\,,
\ee
with
\begin{eqnarray}
        &&\hat{\varepsilon}^\mu(\pm 1) =
                \left[{2\over p^+} \vec \varepsilon_\bot (\pm 1) \cdot
                \vec p_\bot,\,0,\,\vec \varepsilon_\bot (\pm 1)\right],
                \quad \vec \varepsilon_\bot
                (\pm 1)=\mp(1,\pm i)/\sqrt{2}, \nonumber\\
        &&\hat{\varepsilon}^\mu(0)={1\over M_0}\left({-M_0^2+p_\bot^2\over
                p^+},p^+,p_\bot\right).
                \label{polcom}
\end{eqnarray}
  There are several phenomenological light-front wave functions
to describe the possible hadronic structures in the literature. In our work,
we shall use the Gaussian-type and power-law wave functions, given by~\cite{lfda,Hwang2002}
\begin{subequations}
\begin{eqnarray}
\phi_{P}(x,k_{\bot})=\phi_{V}(x,k_{\bot})&=&N\sqrt{\frac{1}{N_c}\frac{dk_{z}}{ dx}} \exp
\left( -\frac{\vec{k}^{2}} {2\omega^{2}}\right) \,,
\label{7a}
 \\
 &=&
 N [x (1-x)]^{1/n} 
\left[ \frac{\omega^{2}} {({\cal A}^{2}+k_{\bot}^{2})+\omega^{2}}\right] \,,
\label{7b}
\\
\phi_{T}(x,k_{\bot})&=&\sqrt{\frac{2}{\omega^{2}}} \phi_{P}(x,k_{\bot}) \,,
\label{7c}
\end{eqnarray}
\end{subequations}
respectively,
where $\omega$ is the scale parameter, $N_c$ is the number of colors, $N = 4 ( \pi/\omega^{2})^\frac{3}{4}$, $\vec k =
(k_{\bot}, k_z)$,  $k_z$ is defined through
\be
x = {E_q+k_z\over
E_q + E_{\bar{Q}}} \,,~~ \ \ 1-x = {E_{\bar{Q}}-k_z \over E_q + E_{\bar{Q}}} \, , ~~\ \
E_i = \sqrt{m_i^2 + \vec k^2} \,
\ee
by
\be
 & &
\ \ k_{z} =\left( x -\frac{1}{2}\right) M_{0}+\frac{m_{q}^{2}-m_{\bar{Q}}^{2}}{%
2M_{0}}~\,,~~ M_0=E_q + E_{\bar{Q}}\, ,
\ee
 $dk_z/ dx = E_q E_{\bar{Q}}/ x(1-x) M_0$, and ${\cal A}=m_{q} x+m_{\bar{Q}}(1-x)$.

The pseudoscalar and vector mesonic decay constants are defined by
\be
\langle 0|A^\mu|P\rangle&=& if_{_P}P^\mu\,,
 \nonumber\\
 \langle 0|V^\mu|V\rangle &=& f_VM_V\epsilon^\mu\,,
 \ee
 where $A^{\mu}=\bar{q}\gamma^{\mu}\gamma_{5}Q$ and $V^{\mu}=\bar{q}\gamma^{\mu}Q$,  respectively.
For  an $^3 P_2$ tensor meson with $J^{PC} = 2^{++}$, the decay constant cannot be produced
through the local $V-A$ and tensor currents.
But, it can be created from the local currents involving covariant derivatives~\cite{hycheng,tdc}:
\be
\langle 0|J_{\mu\nu}|T\rangle=\,f_T M^{2}_{T}\epsilon^{*}_{\mu\nu},
\ee
where
\be
J_{\mu\nu}&=&\frac{i}{2}[\bar{q_1}\gamma_{\mu}\overleftrightarrow{D}_{\nu}q_2
+\bar{q_1}\gamma_{\nu}\overleftrightarrow{D}_{\mu}q_2]  \,.
\ee
and
\be
\overleftrightarrow{D}_{\mu}&=&[\overrightarrow{D}_{\mu}
-\overleftarrow{D}_{\mu}]  \,, \nonumber \\
\overrightarrow{D}_{\mu}&=&\overrightarrow{\partial}_{\mu}-i\frac{g}{2}\lambda^{a}A^{a}_{\mu}\,, \nonumber \\
\overleftarrow{D}_{\mu}&=&\overleftarrow{\partial}_{\mu}+i\frac{g}{2}\lambda^{a}A^{a}_{\mu}\,.
\ee

The polarization tensor $\epsilon_{\mu\nu}$ for a massive spin-two particle can
be constructed out of the polarization vector of a massive vector state~\cite{hycheng,pol},
  given by
\be
\epsilon_{\mu\nu}(\pm2)&=&\epsilon_{\mu}(\pm1)\epsilon_{\nu}(\pm1)  \,, \nonumber \\
\epsilon_{\mu\nu}(\pm1)&=&\sqrt{\frac{1}{2}}[\epsilon_{\mu}(\pm1)\epsilon_{\nu}(0)+\epsilon_{\mu}(0)\epsilon_{\nu}(\pm1)]
 \,, \nonumber \\
\epsilon_{\mu\nu}(0)&=&\sqrt{\frac{1}{6}}[\epsilon_{\mu}(+1)\epsilon_{\nu}(-1)+\epsilon_{\mu}(-1)\epsilon_{\nu}(+1)]
+\sqrt{\frac{2}{3}}\epsilon_{\mu}(0)\epsilon_{\nu}(0) \,.
\label{eps}
\ee

From  the definitions of the meson decay constants, one has
\be
\langle 0|A^\mu|P(p)\rangle &=& -\sqrt{N_c}\int {d^4 k_1 \over{(2 \pi)^4}}
\Lambda_{P}{\rm Tr}\Bigg[\Gamma_{P}
        {i(-\not{\! k_1}+m_{q})\over{k_1^2-m^2_{q}+i\epsilon}}A^{\mu}
       {i(\not{\! p}-\not{\!k_1}+m_{\bar{Q}})\over{(p-k_1)^2-m^2_{\bar{Q}}+i\epsilon}} \Bigg]   \,,
 \nonumber \\
\langle 0|V^\mu|V(p)\rangle &=& -\sqrt{N_c}\int {d^4 k_1 \over{(2 \pi)^4}}
\Lambda_{V}{\rm Tr}\Bigg[\Gamma_{V}
        {i(-\not{\! k_1}+m_{q})\over{k_1^2-m^2_{q}+i\epsilon}}V^{\mu}
       {i(\not{\! p}-\not{\!k_1}+m_{\bar{Q}})\over{(p-k_1)^2-m^2_{\bar{Q}}+i\epsilon}} \Bigg]   \,,
 \nonumber \\
\langle 0|J_{\mu\nu}|T(p)\rangle &=& -\sqrt{N_c}\int {d^4 k_1 \over{(2 \pi)^4}}
\Lambda_{T}{\rm Tr}\Bigg[\Gamma_{T}
        {i(-\not{\! k_1}+m_{q})\over{k_1^2-m^2_{q}+i\epsilon}}J^{\mu\nu}
       {i(\not{\! p}-\not{\!k_1}+m_{\bar{Q}})\over{(p-k_1)^2-m^2_{\bar{Q}}+i\epsilon}} \Bigg] \,,
\label{dc}
\ee
where $\Lambda_{M}$ is a vertex function, related to the
momentum  distrbution amplitude of the $q\bar{Q}$ Fock state.
From  Eqs.~(\ref{n6}) and (\ref{dc}) , we find the vertex function as
\be
\Lambda_{M}&=&\left( \frac{%
k_{1}^{+}k_{2}^{+}}{2[M_{0}^{2}-\left( m_{q}-m_{\bar{Q}} \right) ^{2}]}\right)^{%
\frac{1}{2}} \phi_{M}(x,k_{\bot}) \,,
\ee
where we have  used the light-front variables in Eq.~(\ref{lfvb}).
Then, the explicit expressions of the meson decay constants are given by~\cite{fv,ji}
\be
f_P&=&\,4{\sqrt{3N_c}\over\sqrt{2}}\int {dx\,d^2k_\perp\over 2(2\pi)^3}\,\phi_P(x,
k_\perp)\,{{\cal A}\over\sqrt{{\cal A}^2+k_\perp^2}},
 \nonumber \\
f_V&=&\,4{\sqrt{3N_c}\over\sqrt{2}}\int {dx\,d^2k_\perp\over 2(2\pi)^3}\,\phi_V(x,
k_\perp)\,{1\over\sqrt{{\cal A}^2+k_\perp^2}}\times\bigg\{x(1-x)M^{2}_{0}+m_{q}m_{Q}+k^{2}_{\bot} \nonumber \\
&&~~~~~~~~~+\frac{\cal B}{2W}\bigg[\frac{m_{q}^{2}+k^{2}_{\bot}}{1-x}-\frac{m_{Q}^{2}+k^{2}_{\bot}}{x}
-(1-2x)M^{2}_{0}\bigg]\bigg\}\,,
\nonumber\\
f_T&=&\,4{\sqrt{N_c}\over\sqrt{2}}\int {dx\,d^2k_\perp\over 2(2\pi)^3}\,\phi_T(x,
k_\perp)\,{\frac{1}{x(1-x)\sqrt{{\cal A}^2+k_\perp^2}}}\bigg\{2k^{2}_{\bot}\left[k^{2}_{\bot}(2x-1)^{2}+A^{2} \right]  \nonumber \\
&&~~~~~~~~~+(2x-1)(k^{2}_{\bot}+m_{q}m_{Q})
\left[ (x-1)m_{q}^{2}+xm_{Q}^{2}+(2x-1)k^{2}_{\bot} \right]  \nonumber \\
&&~~~~~~~~~+\frac{1}{2W}\bigg[16 k^{4}_{\bot}x(1-x)(m_{q}+m_{Q}) \nonumber \\
&&~~~~~~~~~+ (1-2 x)^{2} \left(x (m_{q}+m_{Q}) (k^{2}_{\bot}+m_{q}m_{Q})-m_{Q} (k^{2}_{\bot}+m_{q}^{2}) \right)  \nonumber \\
&&~~~~~~~~~  \left(k^{2}_{\bot} (2 x-1)+m_{q}^{2} (x-1)+m_{Q}^{2} x \right)
\bigg]\bigg\}\,,
\ee
where ${\cal A}=m_{q} x+m_{\bar{Q}}(1-x)$, ${\cal B}=m_{q}x-m_{Q}(1-x)$ and $W=M_0+m_q+m_{\bar{Q}}$.

\se{Numerical Results}

\sse{Vector meson decay constants}

\begin{figure}[htbp]
\includegraphics*[width=4in,height=3in,angle=0]{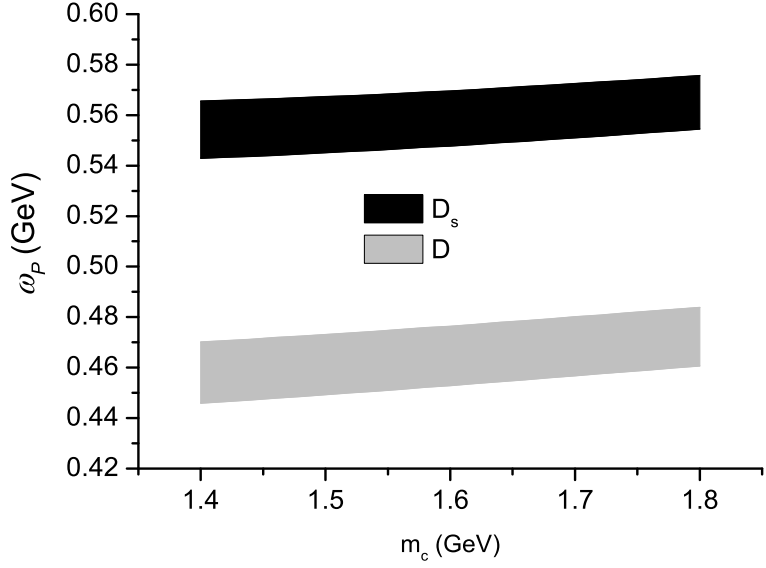}
\caption{Scalar parameters $\omega_P$ ($P=D$ and $D_s$) as functions of  $m_c$ in the LFQM with
 $m_q=0.25$ and $m_s=0.38$ in GeV.}
\label{wd}
\end{figure}

\begin{figure}[htbp]
\includegraphics*[width=4in,height=3in,angle=0]{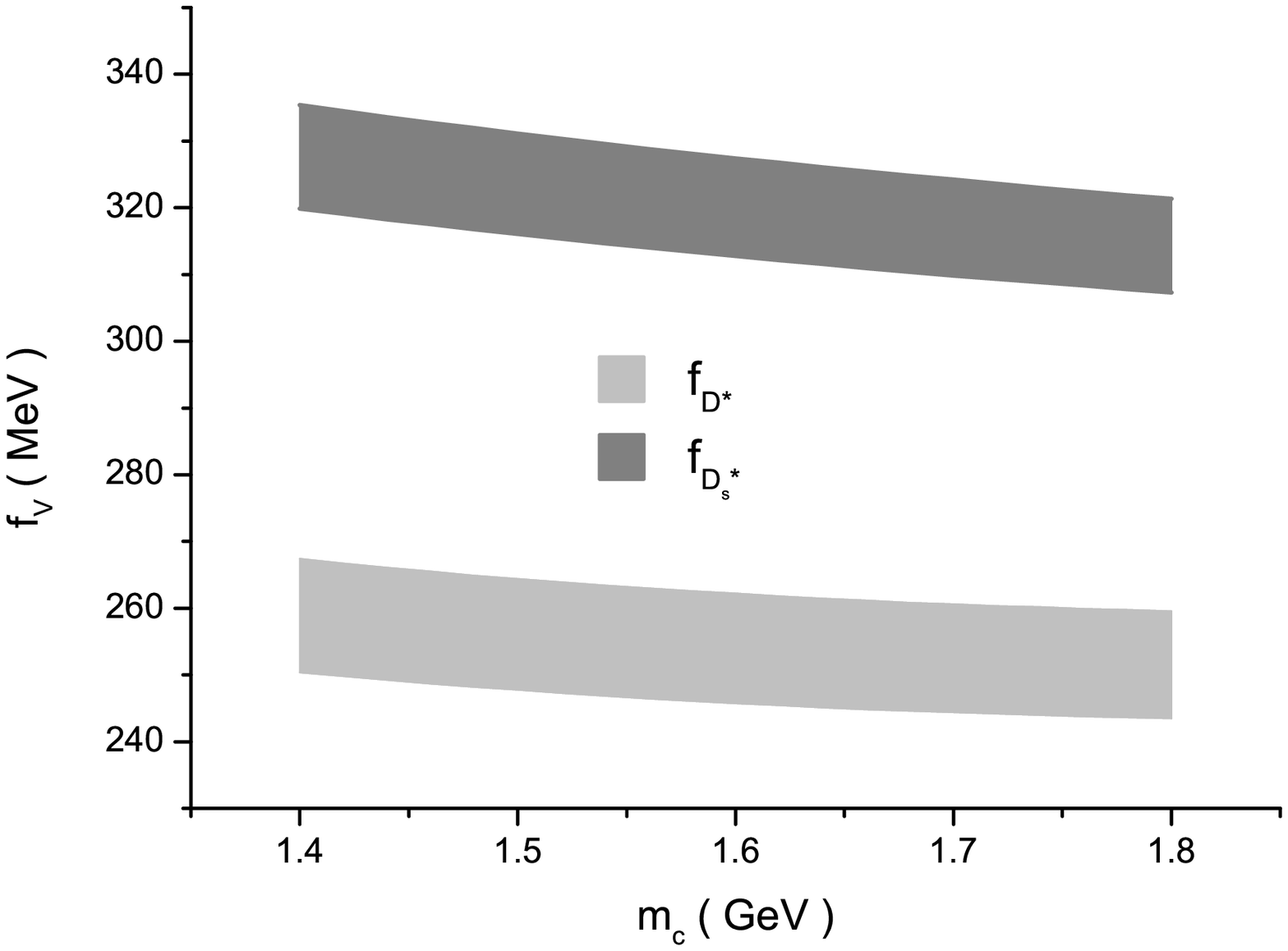}
\caption{$f_{D^{*}}$ and $f_{D^{*}_{s}}$ as  functions of  $m_c$ in the LFQM.}
\label{fvd}
\end{figure}

In the numerical calculation,  we take the known decay constants of
the pseudoscalar mesons ($P$) and  quark masses to
evaluate  the scalar parameters of $\omega_P$. 
For the meson wave functions, we first
use the Gaussian-type wave function in Eq.~(\ref{7a}) and then the power-law one in Eq.~(\ref{7b}). For the latter, we 
only briefly summarize our results.
We start from the decay constants of $f_D$ and $f_{D_{s}}$ from the PDG~\cite{pdg},
given by
\be
f_{D}&=&\,204\pm5\,{\rm
MeV},~~f_{D_{s}}=\,257.5\pm4.6\,{\rm MeV}\,.
\label{fd}
\ee
By using the first equation in Eq.~(17) with the Gaussian-type wave function in Eq.~(\ref{7a}), 
 taking the decay constants in Eq.~(\ref{fd}) and 
inputing the quark masses of $m_u=m_d=0.25$ and $m_s=0.38$  in GeV,
we obtain the parameters $\omega_{D}$ and $\omega_{D_{s}}$
as functions of
the charm quark mass $m_c$, shown in Fig.~\ref{wd}.
In  Fig.~\ref{fvd}, by assuming the parameters of $\omega_{D^{*}}$ and $\omega_{D^{*}_{s}}$ are same as  $\omega_{D}$
and $\omega_{D_{s}}$ with  $m_{u,s}=(0.25, 0.38)$ GeV,
 we plot the decay constants of $f_{D^{*}}$ and $f_{D^{*}_{s}}$ as functions of  $m_c$ in the LFQM, respectively.
 From the figure, we see that the decay constants decrease with $m_c$ but the changes are mild.
Consequently, 
 from Fig.~\ref{fvd} with $m_c=1.5-1.8$ GeV,
 we find
\be
f_{D^*}&=&\,252.0^{+13.8}_{-11.6}\,{\rm
MeV},~~f_{D^{*}_{s}}=\,318.3^{+15.3}_{-12.6}\,{\rm MeV}\,,
\label{fdv}
\ee
which lead to the ratios of the vector and pseudoscalar meson decay constants as
\be
\frac{f_{D^*}}{f_{D}}&=&\,1.232^{+0.074}_{-0.064}\,,~~\frac{f_{D^{*}_{s}}}{f_{D_{s}}}=\,1.236^{+0.063}_{-0.054}\,\,,
\label{rfdv}
\ee
respectively. 
Note that the uncertainties in Eqs.~(\ref{fdv}) come from those of Eq.~(\ref{fd}) and $m_c$, while the errors in
Eqs.~(\ref{rfdv}) result from the combinations of those in Eqs.~(\ref{fd}) and (\ref{fdv}).

\begin{figure}[htbp]
\includegraphics*[width=4in,height=3in,angle=0]{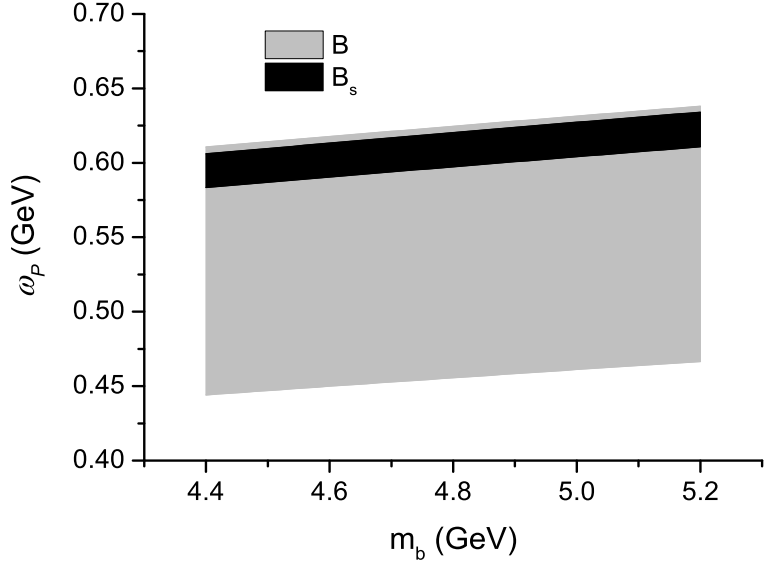}
\caption{ $\omega_P$ ($P=B$ and $B_s$) as functions of $m_b$ in the LFQM with
 $m_q=0.25$ and  $m_s=0.38$ in GeV and the decay constants in Eq.~(\ref{fbb}).}
\label{wb}
\end{figure}
\begin{figure}[htbp]
\includegraphics*[width=4in,height=3in,angle=0]{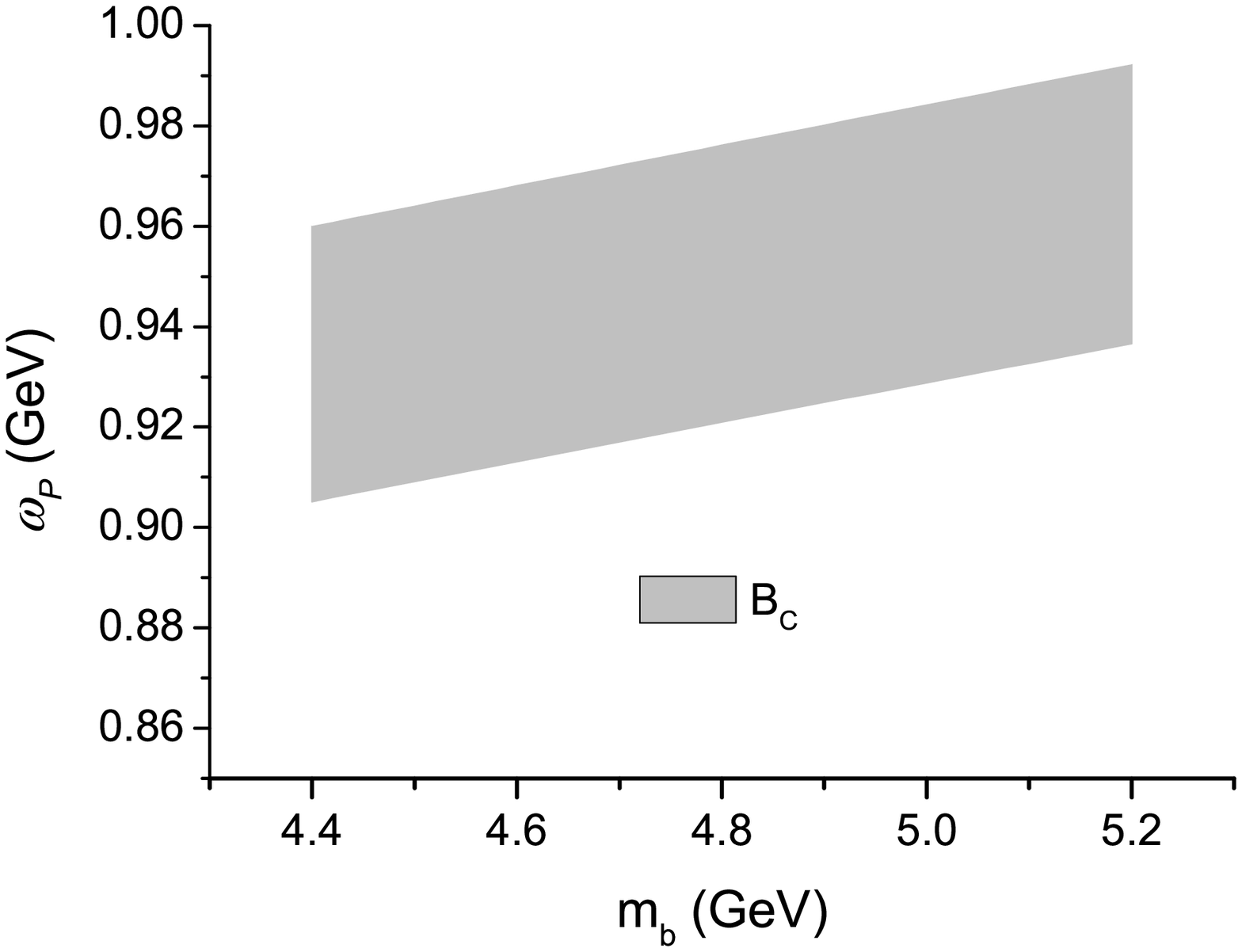}
\caption{Scalar parameters $\omega_P$ ($B_c$) as functions of  $m_b$ in the LFQM with
 $m_c=1.5$ in GeV.}
\label{wBc}
\end{figure}

\begin{figure}[htbp]
\includegraphics*[width=4in,height=3in,angle=0]{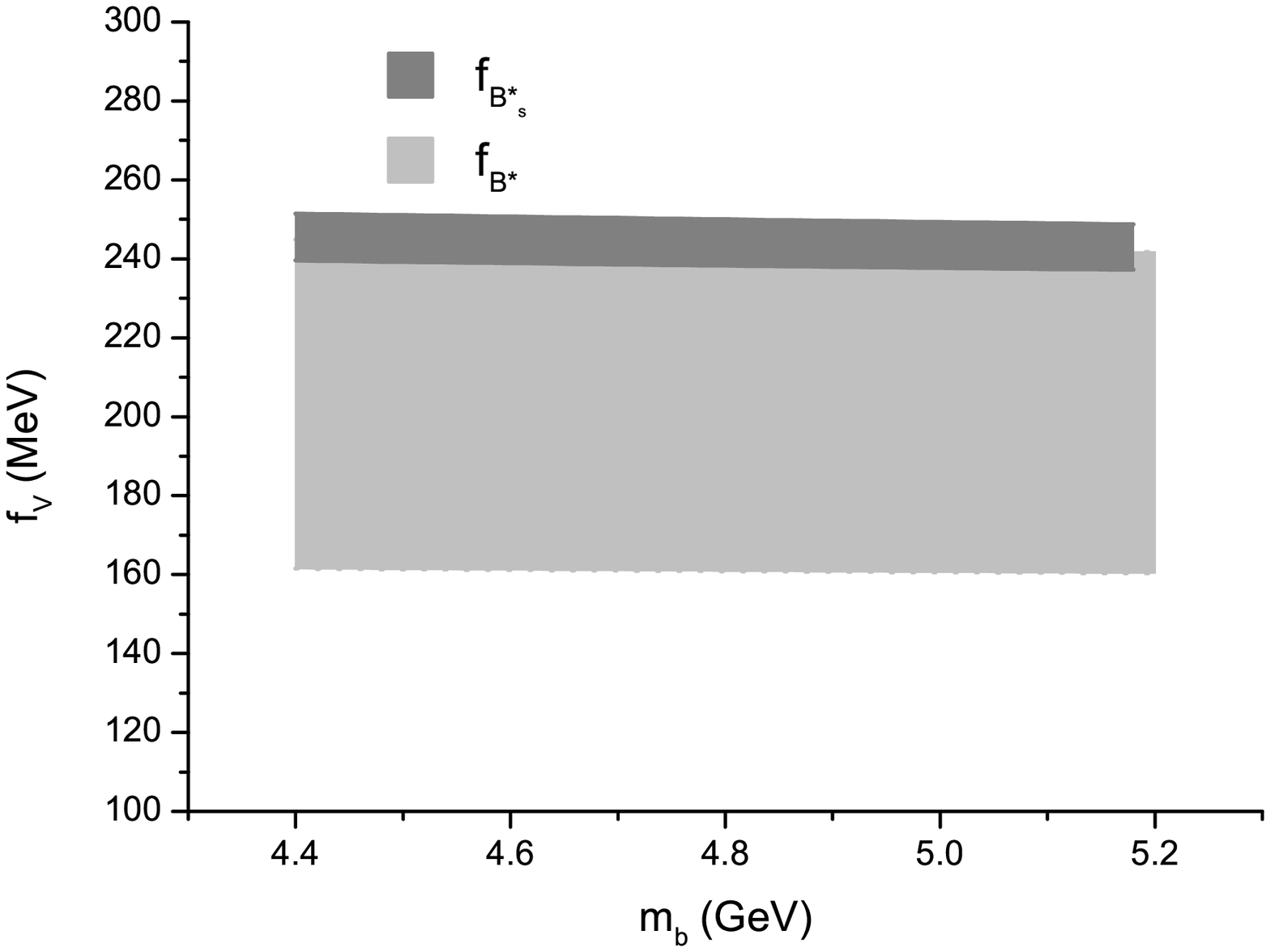}
\caption{$f_{B^{*}}$ and $f_{B^{*}_{s}}$ as  functions of  $m_b$ in the LFQM.
}
\label{fbv}
\end{figure}

\begin{figure}[htbp]
\includegraphics*[width=4in,height=3in,angle=0]{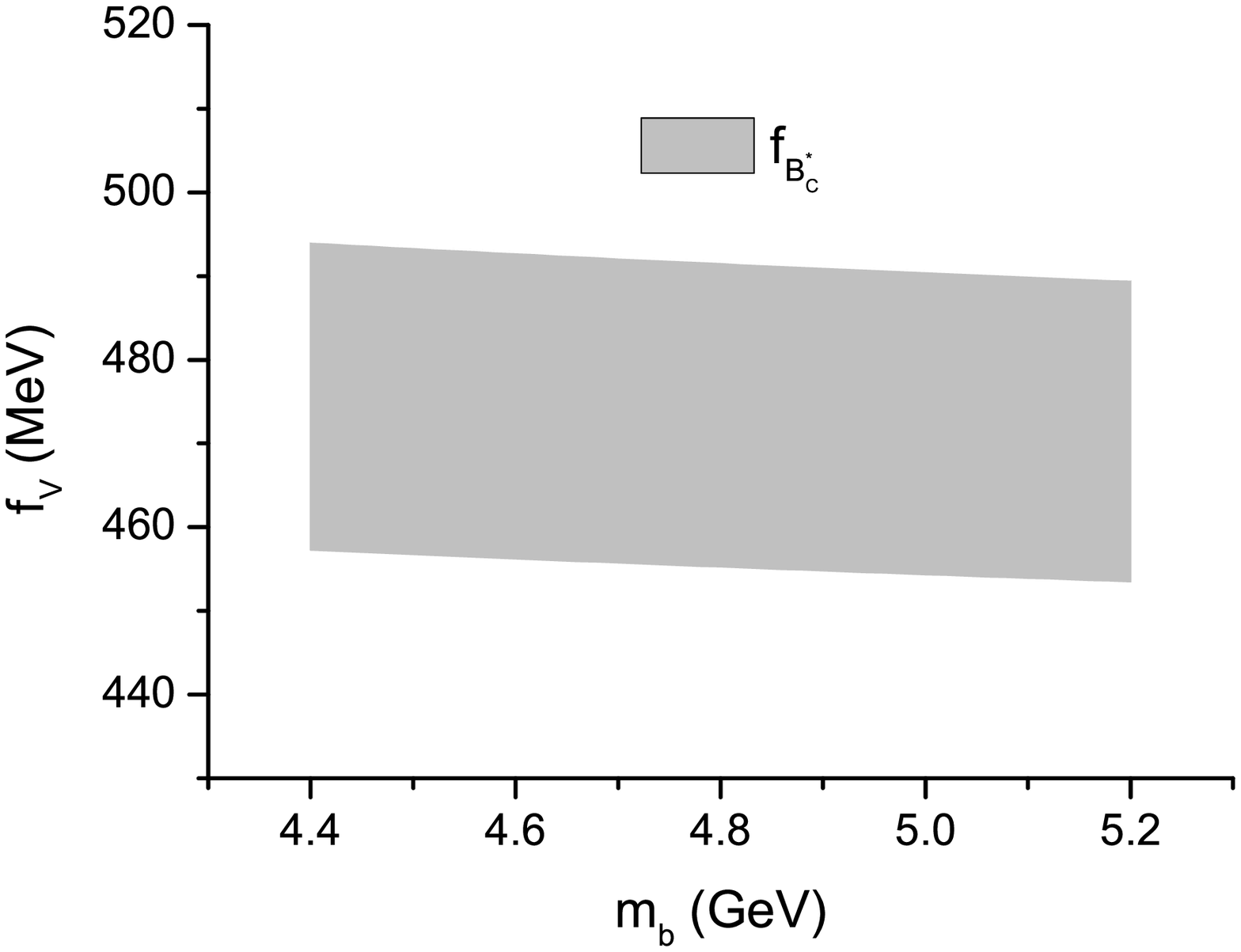}
\caption{$f_{B^{*}_{c}}$ as  functions of  $m_b$ in the LFQM.}
\label{fvBc}
\end{figure}

 From the Belle experimental results~\cite{belle}
and the lattice QCD calculations~\cite{lattice2} of $f_{B}$ ,$f_{B_{s}}$ and $f_{B_{c}}$\cite{Bc},
given by
\be
f_{B}&=&\,185\pm35\,{\rm
MeV},~~f_{B_{s}}=\,224\pm5\,{\rm MeV}\,,~~f_{B_{c}}=\,434\pm15\,{\rm MeV}\,,
\label{fbb}
\ee
 we can fix $\omega_{B}$, $\omega_{B_{s}}$ and $\omega_{B_{c}}$, respectively.
Our results are shown in Figs.~\ref{wb} and \ref{wBc} with $m_{u,s,c}=(0.25, 0.38, 1.5)$ GeV.
In  Figs.~\ref{fbv} and \ref{fvBc},
we present the  decay constants of $f_{B^{*}}$, $f_{B^{*}_{s}}$ and  $f_{B^{*}_{c}}$ as functions of $m_b$ in the LFQM.
Obviously, these decay constants
are insensitive to the change of $m_b$ as seen from the figures.
Similarly, we can derive the ranges of the decay constants $f_{B^{*}}$ and $f_{B^{*}_{s}}$  to be
\be
f_{B^*}&=&\,201.9^{+43.2}_{-41.4}\,{\rm
MeV},~~f_{B^{*}_{s}}=\,244.2\pm7.0\,{\rm MeV}\,,~~f_{B_{c}^{*}}=\,473.4\pm18.2\,~~~{\rm MeV}\,.
\label{fbv1}
\ee
Note that the large error in Eq.~(\ref{fbv1}) for $f_{B^{*}}$ is originated from
the one in
Eq.~(\ref{fbb}) for  $f_{B}$.
Subsequently, 
 we get
the ratios of the vector and pseudoscalar meson decay constants as
\be
\frac{f_{B^*}}{f_{B}}&=&\,1.09^{+0.31}_{-0.30}\,
,~~\frac{f_{B^{*}_{s}}}{f_{B_{s}}}=\,1.09\pm0.04\,\,,~~\frac{f_{B_{c}^{*}}}{f_{B_{c}}}=\,1.09\pm0.06\,.
\label{rfbv}
\ee

In Table~I, we summarize our results  with 
both  Gaussian and  power-law meson wave functions for  the vector meson decay constants.
In the table, we also show 
the  other related theoretical values in the literature~\cite{lcda,lattice,Bc,wangSR,wangSR1,sumrule,sumrule3,sumrule2}.
 From the table, we find that our numerical values with the power-law  wave functions
 are slightly higher than those with the Gaussian ones.
In addition, 
 we can see  that our results for
 $f_{D^{(*)}_{(s)},B^{(*)}_{(s)}}$
are consistent with those from the Lattice QCD~\cite{lattice}
and QCD sum rules (QCDSR) in Refs.~\cite{wangSR,sumrule} but  larger than the ones in Ref.~\cite{sumrule3}.
We note that  $f_{B_{(s)}^*}/f_{B}<1$ in Ref.~\cite{sumrule3}.
For $f_{B^*_c}$, our predicted values are all larger than those in Refs.~\cite{Bc,wangSR1}.
By comparing with Ref.~\cite{lcda}, we see that our predictions
for $f_{D^{*}}$, $f_{D^{*}_{s}}$ and $f_{B^{*}}$
are consistent each other within errors, but those for
$f_{B^{*}_{s}}$ and $f_{B^{*}_{c}}$ are not.
The main reasons for the differences are that the author in Ref.~\cite{lcda} 
used a different set of input parameters such as
quark masses and decay constants of the pseudoscalar mesons.
\begin{table}[htbp]
\caption{ Vector meson decay constants $f_V$ ($V=D^{*}, D^{*}_{s}, B^*, B^{*}_{s,c}$)
in MeV
in this work with (i)  Gaussian and (ii) power-law meson wave functions 
and other theoretical calculations in Refs.~\cite{lcda,lattice,Bc,wangSR,wangSR1,sumrule,sumrule3,sumrule2}.}
 \vskip 0.2in
\label{Table2}
\begin{tabular}{|c||c|c|c|c|c|c|c|} \hline & (i) & (ii)
   & LFQM~\cite{lcda} &  Lattice QCD & QCDSR & QCDSR&QCDSR~\cite{sumrule2}
\\ \hline \hline
$f_{D^{*}}$ & $252.0^{+13.8}_{-11.6}$ & $264.9^{+10.2}_{-9.5}$ &$259.6\pm14.6 $  & $278\pm16 $~\cite{lattice} 
& $263\pm21$~\cite{wangSR}&$252.2\pm22.7$~\cite{sumrule}  &
$242^{+20}_{-12} $
\\ \hline
$f_{D^{*}_{s}}$ & $318.3^{+15.3}_{-12.6}$ &$330.9^{+9.9}_{-9.0}$ & $338.7\pm29.7$  & $ 311\pm9 $~\cite{lattice} 
& $308\pm21$~\cite{wangSR} &$305.5\pm27.3 $~\cite{sumrule}  &
$293^{+19}_{-14} $
\\ \hline
$f_{B^{*}}$ & $201.9^{+43.2}_{-41.4}$ &  $220.2^{+49.1}_{-46.2}$&
$225\pm38$  & $175\pm6$~\cite{Bc} & $213\pm18$~\cite{wangSR}&$181.8\pm13.7 $~\cite{sumrule3}  &
$210^{+10}_{-12} $
\\ \hline
$f_{B^{*}_{s}}$ & $244.2\pm7.0$ &  $265.7\pm8.0$ &
$313\pm67$  & $213\pm7$~\cite{Bc} & $255\pm19$~\cite{wangSR}&$225.6\pm18.5 $~\cite{sumrule3}  &
$251^{+14}_{-16} $
\\ \hline
$f_{B^{*}_{c}}$ & $473.4\pm18.2$ & $487.6\pm19.2$ &
 $387$  & $ 422\pm13$~\cite{Bc} & $384\pm32$~\cite{wangSR1}&$ - $  & $ - $
\\ \hline
\end{tabular}
\end{table}
Finally, we remark that if we take the sharp parameters $\omega_V$ of the vector mesons to be different from  $\omega_P$
of  the pseudoscalar ones, $e.g.$  $\omega_V\sim (1+5\%) \omega_P$, the corresponding vector meson decay constants will
increase about $5\%$ for the same set of input parameters. It is clear that our assumption of $\omega_V\sim \omega_P$
is a consequence of the heavy quark limit, in which $f_P=f_V$ is expected~\cite{Neubert,ChengHQS,HwangHQS}, so that it
may only be applied to the heavy mesons as it is obvious breaking down for the light mesons, such as the case of $\pi$ and $\rho$.

\sse{Tensor meson decay constants}

\begin{figure}[htbp]
\includegraphics*[width=4in,height=3in,angle=0]{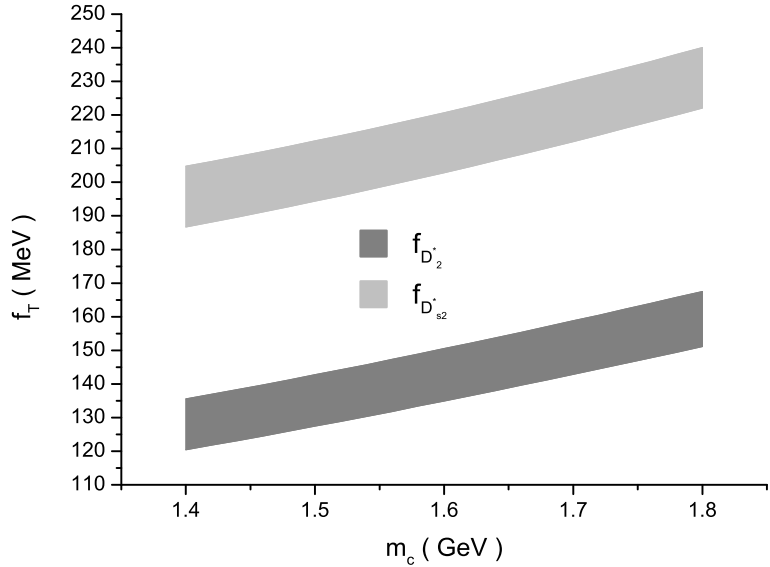}
\caption{$f_{D^*_2}$ and $f_{D^*_{s_2}}$
as  functions of $m_c$ in the LFQM.
}
\label{ftd}
\end{figure}

\begin{figure}[htbp]
\includegraphics*[width=4in,height=3in,angle=0]{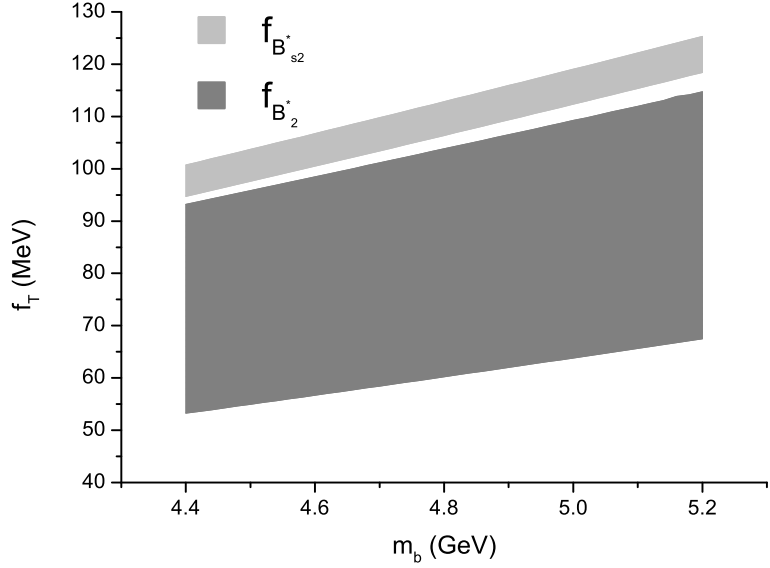}
\caption{$f_{B^*_2}$ and $f_{B^*_{s_2}}$
as  functions of $m_b$ in the LFQM.
}
\label{ftb}
\end{figure}

Similar to the vector meson cases, if we take the parameters of $\omega_T$ are the same as
the corresponding ones for the pseudoscalar mesons,
we may calculate the decay constants of the tensor mesons $D_{2}^{*}$, $D_{s2}^{*}$, $B^{*}_{2}$ and $B^{*}_{s2}$.
In this part of the study, we shall concentrate on the Gaussian-type of the meson wave functions in Eq.~(\ref{7a}).
Note that the relation in Eq.~(\ref{7c}) has been  demonstrated only with the  Gaussian wave functions~\cite{wf3}.
Explicitly, we obtain
\be
f_{D_{2}^{*}}&=&\,143.6^{+24.9}_{-21.8}\,{\rm MeV},~~
f_{D_{s2}^{*}}=\,209.5^{+29.1}_{-24.2}\,{\rm MeV},~~
 \nonumber \\
f_{B^{*}_{2}}&=&\,80.9^{+33.8}_{-27.7}\,{\rm MeV}\,,~~f_{B^{*}_{s2}}=\,109.7^{+15.7}_{-15.0}\,{\rm MeV}\,,
\label{fdt}
\ee
where    $m_{u,s,c,b}=0.25$, $0.38$,  $1.6$ and $4.8$ in GeV have been used to evaluate the center values.
Consequently, we find
the ratios of the two related tensor meson decay constants to be
\be
\frac{f_{D_{s2}^{*}}}{f_{D_{2}^{*}}}&=&\,1.5\pm0.3\,
,~~\frac{f_{B^{*}_{s2}}}{f_{B^{*}_{2}}}=\,1.4^{+0.6}_{-0.5}\,\,.
\label{rfdt}
\ee
In Figs~\ref{ftd} and \ref{ftb}, we show the tensor decay constants of
$D_{2,s_{2}}$
($B_{2,s_{2}}$) as functions of $m_{c(b)}$.
One can see that the decay constants are enhanced if $m_{c(b)}$  increases.

In Table~\ref{Table2}, we  list our results for the tensor meson decay constants
 in the LFQM along with those in QCDSR~\cite{qcdsumrule}.
From the table, we observe  that our predicted value for $D_{s2}^{*}$ is close to that
 in QCDSR, whereas  the other ones are about 20$\%$ smaller.
It is interesting to note that our results in the LFQM can match with those in  QCDSR
if   larger quark masses of $m_{c,b}$ are used.
\begin{table}[htbp]
\caption{ Tensor meson decay constants of  $f_{D_2^*}$, $f_{D_{s2}^{*}}$, $f_{B^{*}_{2}}$ and $f_{B^{*}_{s2}}$  (MeV)
in the LFQM and QCDSR~\cite{qcdsumrule}.}
 \vskip 0.2in
\label{Table2}
\begin{tabular}{|c||c|c|} \hline
   & LFQM &  QCDSR~\cite{qcdsumrule}
\\ \hline \hline
$f_{D_{2}^{*}}$ & $143.6^{+24.9}_{-21.8}$ & $183\pm20 $
\\ \hline
$f_{D_{s2}^{*}}$ & $209.5^{+29.1}_{-24.2}$ & $222\pm22 $
\\ \hline
$f_{B^{*}_{2}}$ & $80.9^{+33.8}_{-27.7}$ & $ 111\pm10 $
\\ \hline
$f_{B^{*}_{s2}}$ & $109.7^{+15.7}_{-15.0}$ & $ 134\pm11 $
\\ \hline
\end{tabular}
\end{table}

\se{Conclusions}

We have studied the decay constants of the heavy vector  ($D^{*}$, $D^{*}_{s}$
$B^{*}$, $B^{*}_{s}$,  $B^{*}_{c}$) and  tensor ($D_{2}^{*}$, $D_{s2}^{*}$, $B^{*}_{2}$, $B^{*}_{s2}$)
 mesons
in the LFQM.
In our study, we have used the known pseudoscalar meson
decay constants of  $f_{D}$, $f_{D_{s}}$, $f_{B}$, $f_{B_s}$ and $f_{B_c}$
and  quark mass $m_{u,d,s}$ and $m_{c(b)}$ as the input parameters to
determine the values of the scale parameters of $\omega_P$ in the light-front wave functions.
By taking $\omega_{D^*_s}$ and $\omega_{B^*_{s,c}}$ in both Gaussian and power-law wave functions being
the same as the corresponding $\omega_{D_{s}}$ and $\omega_{B_{s,c}}$,
we have calculated the decay constants of
the vector $D^*_{(s)}$ and $B^*_{(s,c)}$ mesons, respectively.
Explicitly, we have found that
$f_{D^{*}, D^{*}_{s}, B^{*},B^{*}_s,B^{*}_c} =
(252.0^{+13.8}_{-11.6}$, $318.3^{+15.3}_{-12.6}$ , $201.9^{+43.2}_{-41.4}$, $244.2\pm7.0$, $473.4\pm18.2$) and
$(264.9^{+10.2}_{-9.5}$, $330.9^{+9.9}_{-9.0}$, $220.2^{+49.1}_{-46.2}$, $265.7\pm8.0$, $487.6\pm19.2$) MeV
with  Gaussian and power-law wave functions,
  respectively.
Similarly, we have obtained
$f_{D_{2}^{*},D_{s2}^{*},B^{*}_{2},B^{*}_{s2}}$=
 ($143.6^{+24.9}_{-21.8}$, $209.5^{+29.1}_{-24.2}$,
 $80.9^{+33.8}_{-27.7}$, $109.7^{+15.7}_{-15.0}$) MeV
 with only Gaussian wave functions.

\section{Acknowledgments}
The work was supported in part by National Center for Theoretical Sciences, National Science
Council (NSC-101-2112-M-007-006-MY3 and NSC-102-2112-M-471-001-MY3) and
MoST (MoST-104-2112-M-007-003-MY3).

\end{document}